\documentclass[12pt,preprint]{aastex}

\def\vec#1{\mbox{\boldmath $#1$}}

\slugcomment{Not to appear in Nonlearned J., 45.}

\shorttitle{3D MHD Dynamics of the X2.2-Class Flare on 2011 Feb./ 15}
\shortauthors{Inoue et al.}

\begin{document}


\title{Magnetohydrodynamic Simulation of the X2.2 Solar Flare \\ 
       on 2011 February 15: II. Dynamics Connecting   
       the Solar Flare and the Coronal Mass Ejection}


\author{S.\  Inoue}
\affil{ School of Space Research, Kyung Hee University, Yongin 446-701, Korea}
\email{inosato@khu.ac.kr}

\author{K.\ Hayashi,}
\affil{ Key Laboratory of Solar Activity, National Astronomical Observatories of 
        China,\\  Chinese Academy of Sciences, Beijing, 100012, China}
\altaffiltext{1}{W.\ W.\ Hansen Experimental Physics Laboratory, Stanford 
             University Stanford, CA 94305 USA}
\altaffilmark{1}

\author{T.\ Magara}
\affil{ School of Space Research, Kyung Hee University, Yongin 446-701, Korea}

\author{G.\  S.\  Choe}
\affil{ School of Space Research, Kyung Hee University, Yongin 446-701, Korea}

\and 

\author{Y.\  D.\  Park}
\affil{ Korean Astronomy and Space Science Institute, Daejeon 305-348, Korea}

  \begin{abstract}
  We clarify a relationship of the dynamics of a solar flare and a growing 
  Coronal Mass Ejection (CME) by investigating the dynamics of magnetic fields 
  during the X2.2-class flare taking place in the solar active region 11158 
  on 2011 February 15, based on simulation results obtained from 
  \cite{2014ApJ...788..182I}. We found that the strongly twisted lines formed 
  through the tether-cutting reconnection in the twisted lines of a nonlinear 
  force-free field (NLFFF) can break the force balance within the magnetic field, 
  resulting in their launch from the solar surface. We further discover that a 
  large-scale flux tube is formed during the eruption as a result of the 
  tether-cutting reconnection between the eruptive strongly twisted lines and 
  these ambient weakly twisted lines. Then the newly formed large flux tube 
  exceeds the critical height of the torus instability. The tether-cutting 
  reconnection thus plays an important role in the triggering a CME. Furthermore,
  we found that the tangential fields at the solar surface illustrate different 
  phases in the formation of the flux tube and its ascending phase over the 
  threshold of the torus instability. We will discuss about these dynamics in 
  detail.
  \end{abstract}

  \section{Introduction}
   Solar flares and coronal mass ejections (CMEs) are powerful explosions of 
   the solar coronal plasma that are widely considered to be sudden liberation 
   of the free magnetic energy in the solar atmosphere
   (\citealt{2000JGR...10523153F}; \citealt{2002A&ARv..10..313P}). Our 
   understanding of them is important so that we can not only clarify the 
   nonlinear dynamics of the solar plasma, {\it i.e.}, the store-and-release 
   processes for magnetic energy and heilicty but also establish the space 
   weather forecast. Since \citealt{1859MNRAS..20...13C} discovered the solar 
   flare, many observational, theoretical, and numerical studies have been 
   conducted in order to clarify the properties of solar flares and their 
   relationship with CMEs
   (\citealt{2008LRSP....5....1B}; 
    \citealt{2011LRSP....8....6S}).

   \cite{1964CSHKP}, \cite{1966Natur.211..695S}, \cite{1974SoPh...34..323H} and 
   \cite{1976SoPh...50...85K} constructed a flare model based on magnetic 
   reconnection, which well explained observations of solar flares made at 
   multiple wavelengths. This model is named the CSHKP model and is today 
   considered a standard flare model. Since this model was formed, solar physics 
   satellites have provided countless data along with images of the solar flares 
   and CMEs, which are further clarifying their behaviors. In particular, the 
   {\it Yohkoh} satellite convinced us of a flare scenario based on the magnetic 
   reconnection (\citealt{1992PASJ...44L..63T}; \citealt{1994Natur.371..495M}). 
   In addition, \cite{2001ApJ...546L..69Y} discovered clear evidence of 
   reconnection inflows associated with the solar flare from the observations of 
   {\it Yohkoh}/SXT and {\it SOHO}/EIT. All of these observations built up a 
   foundation of the reconnection model for the solar flare as well as support 
   the classical CSHKP flare model. These observations and theoretical models 
   clarified the properties of solar flares and helped us understand flare 
   physics in a two dimensional regime. On the other hand, our understanding of 
   the dynamics of solar flares is not fully developed in a three dimensional 
   (3D) regime. 

    Recently, an innovative development of a computer allows us to begin closing 
   this deficiency in our knowledge by performing a 3D magnetohydrodynamic (MHD) 
   simulation to explore solar flare dynamics in 3D space. 
   \cite{1999ApJ...510..485A} performed 3D simulation and propose a 
   magnetic breakout model showing an onset of the CME by removing a overly field
   lines above the core field, through a reconnection. The model has been  
   extended and high resolved simulations are recently performed ({\it e.g.}, 
   \citealt{2008ApJ...683.1192L}, \citealt{2012ApJ...760...81K}).
   \cite{2000ApJ...529L..49A} and \cite{2003ApJ...585.1073A} successfully showed 
   the formation of a flux tube and its eruption via a twist and converging 
   motion on the photosphere from an  initial state which starts from a potential
   field. \cite{2005ApJ...630L..97T} confirmed that the helical kink instability 
   well explains the initial processes of a CME started with anchoring a flux 
   tube in the photosphere and embedding it in the solar corona. They further 
   pointed out that the decrease in the overlying field is a key factor 
   determining whether the resulting eruption fails or is ejective. 
   \cite{1990JGR....9511919F} constructed the flux tube model in 2D space and 
   pointed out the eruption via the loss of equilibrium. Following 
   \cite{1990JGR....9511919F}, \cite{2006ApJ...645..742I} applied a simple 
   straight flux tube in the 3D solar corona; however, \cite{2006ApJ...645..742I}
   indicated that the flux tube exists in an unstable state against the kink 
   instability before it loses its equilibrium. They further suggested an 
   existence of a critical height deciding whether the eruption is full or 
   failed. \citealt{2010ApJ...708..314A} and \cite{2010ApJ...719..728F} 
   well-reproduced the initiation of a CME driven by the torus instability
   (\citealt{2006PhRvL..96y5002K}). \cite{2012A&A...543A.110A} brought an 
   interpretation of the CSHKP model into 3D space, explaining the 3D effect, 
   {\it e.g.}, the strong-to-weak shear transition in the post-flare loop which 
   is not seen in the classical CSHKP model. \citealt{2012ApJ...760...31K} 
   explained that two different types of emerging small fluxes play a role in 
   the triggering mechanisms for the solar flare and CMEs. Recently, a more 
   realistic situation is constructed; where the initial condition mimics a real 
   situation or is applied to a potential field extrapolated from the vector 
   field, enabling us to compare 3D simulations with real observations
   ({\it e.g.,} \citealt{2012ApJ...758..117Z}; \citealt{2014ApJ...788...85V}). 
   More recently, \cite{2013ApJ...773...21A}, \cite{2013PASJ...65L...5M} and 
   \cite{2014ApJ...787...46L} are trying to understand the onset mechanism from 
   their results on the flux emergence simulation without constructing a model. 

   These 3D simulations show complicated and specific nonlinear dynamics only 
   observable in 3D space. In addition, they interpolate well between 3D 
   dynamics and the observations or models constructed in 2D space. Recently, a 
   solar optical telescope (\citealt{2008SoPh..249..167T}) and the Helioseismic 
   and Magnetic Imager (\citealt{2012SoPh..275..207S};
   \citealt{2014SoPh..289.3483H}) on board the latest solar physics satellite 
   {\it Hinode} (\citealt{2007SoPh..243....3K}) and 
   {\it Solar Dynamics observatory (SDO:} \citealt{2012SoPh..275....3P}) 
   have provided vector magnetic fields in high temporal and spacial resolutions.
   These vector fields allow us to reconstruct the 3D coronal magnetic field 
   under a nonlinear force-free field (NLFFF) approximation
   (\citealt{2012LRSP....9....5W}; \citealt{2013SoPh..288..481R}) which takes 
   into account even the tangential components of magnetic fields of the vector 
   magnetic fields and is much different from the potential field extrapolated 
   only from the normal component of the vector magnetic fields. The NLFFF 
   exhibits the strong sheared magnetic fields appearing above the polarity 
   inversion line (PIL) prior to the solar flares and CMEs ({\it e.g.} 
   \citealt{2004A&A...425..345R} or \citealt{2010ApJ...715.1566C}), as observed 
   by ground and space observations. These are not shown in the potential field.
   On the other hand, it shows the relaxed magnetic field lines after the flare 
   (\citealt{2008ApJ...675.1637S};
   \citealt{2012ApJ...760...17I};
   \citealt{2014JGRA..119.3286H}).
   Some papers further clear the temporal evolution of the magnetic field in 
   the active region throughout the flare event
   (\citealt{2008A&A...484..495T};
   \citealt{2011ApJ...738..161I};
   \citealt{2012ApJ...748...77S}; 
   \citealt{2013ApJ...770...79I};
   \citealt{2014ApJ...780...55J};
   \citealt{2014ApJ...789...93C}).  
   However, because these NLFFFs are constructed in the steady sate, they never 
   reveal the dynamics of the solar flares. Therefore, the MHD simulation which
   takes into account the NLFFF, {\it i.e.}, data-constrained simulation, are 
   taking a new approach
   (\citealt{2013ApJ...771L..30J}; 
    \citealt{2013ApJ...779..129K};
    \citealt{2013A&A...554A..77P}).
   
   More recently, \cite{2014ApJ...788..182I} performed the MHD simulation 
   using the NLFFF as an initial condition in order to elucidate the dynamics   
   of the solar flare. This NLFFF is reconstructed 2 hours approximately before 
   the X2.2-class flare in AR11158. The AR11158, which consists of a 
   complicated quadrupole structure, produced several M- and X-class flares in  
   2011 February (\citealt{2011ApJ...738..167S}; \citealt{2012ApJ...748...77S};
   \citealt{2013ApJ...770...79I}) and Hinode and SDO satellites provided rich 
   data observed in multiple wavelengths over the course of the flare events. 
   \cite{2014ApJ...788..182I} showed that the dynamics obtained from our MHD 
   simulation successfully reproduced observed phenomena such as the 
   distribution of the two-ribbon flare and the field lines structure as seen 
   in the extreme ultraviolet (EUV) post-flare image taken by Atmospheric 
   Imaging Assembly (AIA; \citealt{2012SoPh..275...17L}) which was on board 
   {\it SDO}. However, the detailed dynamics of this event are not yet clear.
  
   In this study, we explore the detailed dynamics of the magnetic field during 
   the X2.2 solar flare taking place in AR 11158, based on the simulation 
   results obtained by \cite{2014ApJ...788..182I}. In particular, we clarify the 
   dynamics connecting the solar flare and the CME, {\it i.e.} the geometry and 
   evolution encompassing the region of the flare and CME initiation. This paper
   is constructed as follows. Our observational data and numerical method are 
   described in Section 2. Our results and discussions are presented in Sections
   3 and 4. Our conclusion is summarized in Section 5. 
          
  \section{NLFFF Extrapolation and MHD Simulation Methods}
  \subsection{Numerical Method and Observations}
   All of the methods for the NLFFF and MHD simulation following  
  \cite{2014ApJ...788..182I}. Therefore, here we briefly provide an overview 
  of the numerical methods and observations presented in that work to aid in 
  readers understanding. The NLFFF and MHD simulations are based upon the 
  following equations; 
  
  \begin{equation}
  \rho = |\vec{B}|,
  \label{eq_den_NLFFF}
  \end{equation}

  \begin{equation}
  \frac{\partial \vec{v}}{\partial t}
                        = - (\vec{v}\cdot\vec{\nabla})\vec{v}
                          + \frac{1}{\rho} \vec{J}\times\vec{B}
                          + \nu\vec{\nabla}^{2}\vec{v},
  \label{eq_motion}
  \end{equation}

  \begin{equation}
  \frac{\partial \vec{B}}{\partial t}
                        =  \vec{\nabla}\times(\vec{v}\times\vec{B}
                        -  \eta_{i}\vec{J})
                        -  \vec{\nabla}\phi,
  \label{induc_eq}
  \end{equation}

  \begin{equation}
  \vec{J} = \vec{\nabla}\times\vec{B},
  \end{equation}

  \begin{equation}
  \frac{\partial \phi}{\partial t} + c^2_{h}\vec{\nabla}\cdot\vec{B}
    = -\frac{c^2_{h}}{c^2_{p}}\phi,
  \label{div_eq}
  \end{equation}
  where the subscript $i$ of $\eta$ corresponds to 'NLFFF' or 'MHD'. The length, 
  magnetic field, density, velocity, time, and electric current density are 
  normalized by   
  $L^{*}$ = 216   Mm,  
  $B^{*}$ = 2500 G, 
  $\rho^{*}$ = $|B^{*}|$,
  $V_{A}^{*}\equiv B^{*}/(\mu_{0}\rho^{*})^{1/2}$,    
  where $\mu_0$ is the magnetic permeability,
  $\tau_{A}^{*}\equiv L^{*}/V_{A}^{*}$, and     
  $J^{*}=B^{*}/\mu_{0} L^{*}$,  
  respectively. The non-dimensional viscosity $\nu$ is set as a constant  
  $(1.0\times 10^{-3})$, and the coefficients $c_h^2$, $c_p^2$ in 
  equation (\ref{div_eq}) also fix the constant values, 0.04 and 0.1, 
  respectively. Although the density is given as a model like
  \cite{1996ApJ...466L..39A}, we previously discussed its validity in 
  \cite{2014ApJ...788..182I}.

  The vector magnetic field 
  \footnote{Data is available in http://jsoc.stanford.edu/jsocwiki/ReleaseNotes 
  or http://jsoc.stanford.edu/new/HMI/HARPS.html and 
  \cite{2014SoPh..289.3549B}} that we used as the boundary condition is 
  observed at 00:00 UT on February 15, approximately 2 hours before the 
  X2.2-class flare was detected by HMI/{\it SDO}. It covers a 216 $\times$ 216 
  (Mm$^2$) region, divided into a 600 $\times$ 600 grid. It is obtained using 
  the very fast inversion of the Stokes vector (VFISV) algorithm 
  (\citealt{2011SoPh..273..267B}) based on the Milne$-$Eddington
  approximation. A minimum energy method (\citealt{1994SoPh..155..235M};
  \citealt{2009SoPh..260...83L}) was used to resolve the 180$^\circ $ ambiguity 
  in the azimuth angle of the magnetic field. In this study, the vector field 
  is preprocessed in accordance with \cite{2006SoPh..233..215W}.

   A numerical box with dimensions of 216 $\times$ 216 $\times$ 216 (Mm$^3$) 
  is given by 1 $\times$ 1 $\times$ 1 as a non-dimensional value. 2 $\times$ 2 
  binning of the original data yields grid numbers which are assigned as 
  300 $\times$ 300 $\times$ 300. The detailed numerical scheme is available  
  in \cite{2013ApJ...770...79I} or \cite{2014ApJ...780..101I}.

  \subsection{NLFFF Extrapolation}
   We first perform an NLFFF extrapolation to obtain the 3D structure of the 
  magnetic field prior to the X2.2-class flare, and to understand its physical 
  properties relating to stability; we apply the MHD simulation as an initial 
  condition. This is based on the MHD relaxation method developed by 
  \cite{2014ApJ...780..101I} which shows the detailed algorithm. The initial 
  condition is given by a potential field obtained from the Green function 
  method (\citealt{1982SoPh...76..301S}). At the boundaries, the three 
  components of the magnetic field are fixed; in particular, the three observed 
  components of the magnetic fields are set at the bottom boundary and their 
  velocities are set as zero. The Neumann boundary condition is imposed on 
  $\phi$. A formulation of $\eta$ is given by
  \begin{equation}
  \eta_{nlfff} = 5 \eta_0 + \eta_1 \frac
                {|\vec{J}\times\vec{B}||\vec{v}|^2}{\vec{|B|^2}},
  \label{eta_nlfff}
  \end{equation}
  where $\eta_0$=$1.0 \times 10^{-5}$ and $\eta_1$=$1.0 \times 10^{-3}$.
  The NLFFF is performed in the center area of AR11158 to avoid any effects 
  from the inconsistent force-free $\alpha$, and outside it is fixed by the 
  potential field (see Figure 1(a) in \citealt{2014ApJ...788..182I}). 
  This calculation corresponds to Run A in Table 1 as shown in 
  \cite{2014ApJ...788..182I}. Figure \ref{f1}(a) shows a distribution of the 
  magnetic twist and 3D field lines structure in the central area of AR11158. 
  Magnetic twist is defined by
  \begin{equation}
  T_n = \frac{1}{4\pi}\int \frac{J_{||}}{|\vec{B}|} dl,
  \label{Tn_Eq}
  \end{equation}
  where $J_{||} = \vec{J}\cdot\vec{B}/|\vec{B}|$ and $dl$ is a line element
  (\citealt{2006JPhA...39.8321B}). We found that the strong magnetic twist of 
  the NLFFF is distributed in the range from a half-turn to one-turn twist. 
  \cite{2014ApJ...788..182I} found that the NLFFF exists in a stable state not 
  only against an ideal MHD instability, such as a kink or torus instability, 
  but also against even a small perturbation. In other words, this NLFFF needs 
  to be excited into a higher energy level to release the free magnetic energy 
  required to produce the flare. 
      
  \subsection{MHD Relaxation and Simulation}
\begin{figure}
  \epsscale{.9}
  \plotone{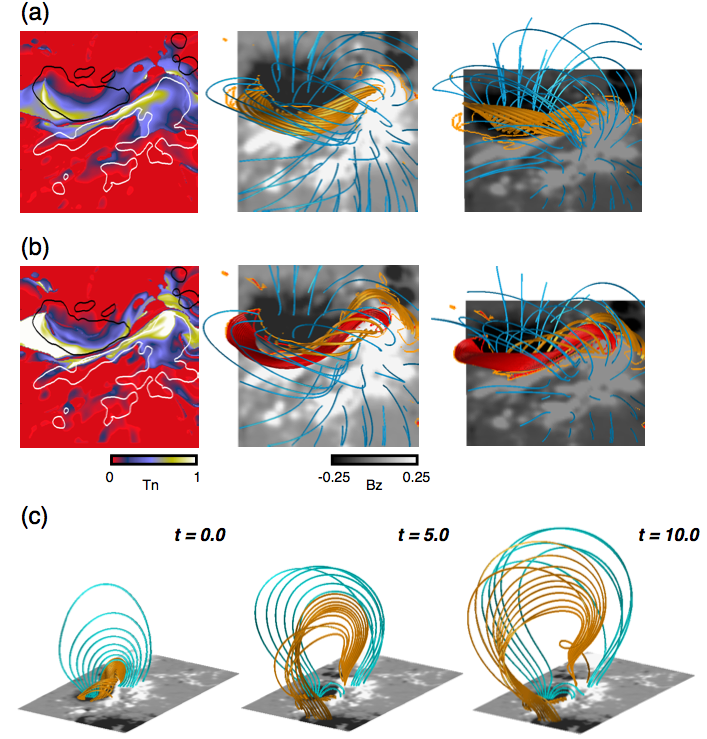}
  \caption{
           (a) The left panel is a distribution of the magnetic twist of the 
               NLFFF 2 hours approximately before the X2.2-class flare, with 
               contours of $B_z$ where black and white contours correspond to 
               $B_z$=0.25(625 G), and -0.25, respectively. The field lines 
               begin in the region (0.35$L_0$ $<$x$<$ 0.65$L_0$ and 
               0.4$L_0$ $<$y$<$ 0.7$L_0$, where $L_0$ is 216 Mm ). Then the 
               magnetic twist is mapped, where the value of $|B_z|$ is more than 
               0.01 (=25 G). The middle panel shows the field line structure. 
               The orange curves show twisted lines with more than a half-turn 
               twist while the blue lines have less than a half-turn twist. The 
               Right panel also represents the 3D structure of the field lines 
               from another angle. 
           (b) The twist distribution and 3D field line structure after the MHD 
               relaxation process. The figure format is the same as (a), except 
               the red lines show the strongly twisted lines with more than a 
               one-turn twist.
           (c) The temporal evolution of the magnetic field obtained from the MHD
               simulation. The orange and blue lines have twisted lines more and 
               less than a half-turn twist, respectively, at an initial time 
               $t$=0.0.    
           }
  \label{f1}
  \end{figure}

  We further perform the MHD relaxation using the NLFFF as an initial condition
  to excite it into higher energy level. We employ the zero-$\beta$ MHD 
  equations which are the same as those used in the NLFFF calculation, except 
  the velocity limiter is released (see \citealt{2014ApJ...780..101I}) and the 
  resistivity formulation is replaced by the anomalous one as follows,
  \begin{equation}
  \eta_{MHD} (t) = \left\{
         \begin{array}{ll}
         \eta_{0} & \quad \mbox{$J<j_{c}$}, \\
         \eta_{0}+\eta_{2} 
              (\frac{J-j_{c}}{j_{c}})^{2} & \quad \mbox{$J>j_{c}$},
         \end{array}\right.
   \label{ano_res}
   \end{equation}
  where $\eta_{2}=5.0\times 10^{-4}$, and $j_{c}$ is the threshold current, set 
  to 30 in this study. The boundary condition is same as a case of the 
  NLFFF. Because the anomalous resistivity can enhance a reconnection in the 
  strong current region between the twisted lines formed in the NLFFF, we can 
  expect it to form stronger twisted lines. This calculation corresponds to 
  Run C in Table 1 as shown in \cite{2014ApJ...788..182I}. Figures \ref{f1}(b) 
  shows the distribution of the magnetic twist and 3D field lines structure after
  the MHD relaxation process, these are same format with Figure \ref{f1}(a). We 
  found that the strongly twisted lines with more than a one-turn twist are 
  formed after the MHD relaxation process, which are not shown in the original 
  NLFFF, as shown in Figure \ref{f1}(a). The formation of these strongly twisted 
  lines is clearly due to the anomalous resistivity, which is reminiscent of the 
  tether-cutting reconnection.

   Finally, we perform the MHD simulation using the magnetic field shown in 
  Figure \ref{f1}(b) obtained after the MHD relaxation process. At the 
  boundaries, the tangential components of the magnetic field are released to 
  interact with an induction equation consistently, while the normal component 
  of the field remains constant. Note that in this case, the tangential 
  components gradually go back toward the state of the potential field 
  (or another less twisted state) because the NLFFF cannot maintain the 
  force-free state completely. The velocity and $\phi$ are treated in the same 
  manner for all boundaries as were the NLFFF calculation and time-relaxed 
  simulation. This calculation corresponds to Run D in Table 1 as shown in 
  \cite{2014ApJ...788..182I}. The magnetic field shown in Figure \ref{f1}(b) 
  is already in a non-equilibrium state;therefore, it illustrates the dramatic 
  dynamics shown in Figure \ref{f1}(c). An overview of these is was summarized 
  in \cite{2014ApJ...788..182I}.

  \section{Results}
  \cite{2014ApJ...788..182I} presented the overview of the dynamics and compared 
  these with some observations. In this paper, we study the more detailed 3D 
  dynamics, in particular, focusing on the formation of the large flux tube, 
  {\it i.e.}, the growing process of the CME, and other phenomena associated 
  with the reconnection, eventually understanding a relationship flares and 
  CMEs.

  \subsection{Temporal Evolution of the Magnetic Twist}
    In general, 3D dynamics obtained from the MHD simulation show a complicated
   behavior, so it is not easy to extract essential components from them. 
  However, the magnetic twist defined in equation (\ref{Tn_Eq}) is one solid 
  tool that we can use to extract these components, {\it e.g.}, a dynamics of 
  flux tube and the reconnection associated with the formation and dynamics of 
  the flux tube. Because the flux tube consists of a bundle of field lines with 
  a strong magnetic twist, we can trace the temporal evolution of the flux tube 
  by tracing that of the magnetic twist. Furthermore, since the magnetic twist 
  is a magnetic helicity generated by a toroidal current in the flux tube 
  (\citealt{2006JPhA...39.8321B}; 
   \citealt{2010A&A...516A..49T};
   \citealt{2012ApJ...760...17I}
  ), the magnetic twist would help us understand the reconnection dynamics 
  in terms of conservation law of the magnetic helicity.

   We first show in Figure \ref{f2} the temporal evolution of the magnetic 
  twist during the flare obtained from the MHD simulation, mapped on the solar 
  surface where $T_n=1.0$ and $T_n$=0.5 are plotted. The locations 
  surrounded by the contour of $T_n$=1.0 and 0.5 correspond to where the base
  of the flux tube exists. We clearly see that the locations of the twisted 
  flux tube are not fixed;rather, they move in a course. At positive polarity, 
  the region occupied by the strong magnetic twist is moving southward. The 
  region located at the negative polarity is shrinking in an early phase and 
  then seems to maintain its size in a later phase. These are not simple 
  pictures that we can image easily, particularly since they are occurring in 
  3D space. Furthermore, although the twist is widely distributed in the area 
  close to the polarity inversion line (PIL) at initial time, it seems to be 
  concentrated in the area away from the PIL at $t$=12.5. In other words, most 
  of the strongly twisted lines disappear close to the PIL. The reason for this 
  is that the post-flare loops are formed through reconnection during the flare,
  as shown in Figure 8(a) of \cite{2014ApJ...788..182I}.
 
  \subsection{Formation of the Large Flux Tube}
\begin{figure}
  \epsscale{1.}
  \plotone{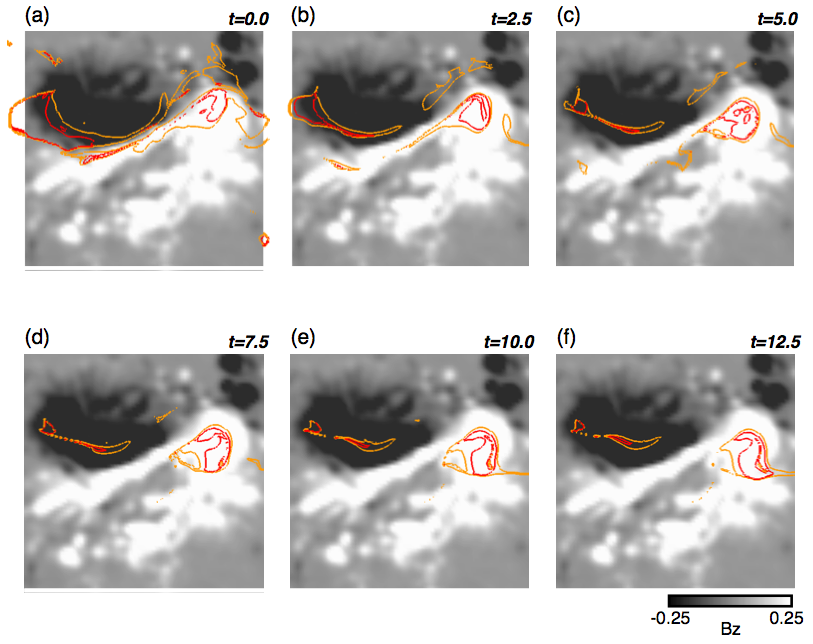}
  \caption{
          Temporal evolution of the magnetic twist mapped on the solar surface,
          focusing on the central region of AR11158. The $B_z$ distribution 
          is represented in gray scale. The red and orange lines are contours of 
          the magnetic twist $T_n=1.0$ and $T_n=0.5$, respectively. Note that 
          the inside regions surrounded by the red and orange lines are occupied 
          by the strongly twisted lines. 
          }
  \label{f2}
  \end{figure}

  We see the temporal evolution of the magnetic field lines in Figure \ref{f3},  
  where we focus on the strongly twisted lines in red which have more than 
  one-turn twist and those in blue which have less than a one-turn twist (since 
  most parts are less than half-turn, hereafter we call them weakly twisted 
  lines). These weakly twisted lines are traced from the location close to  
  where the strongly twisted lines exit. Although the strongly twisted lines 
  with more than a one-turn twist launch away from the solar surface, they split
  into two parts in an early phase, $t$=2.5; one footpoint of some field 
  lines is rooted in the negative polarity at southeast area which locates away 
  from the central area. That is why we see that the strongly twisted region 
  occupied on central negative polarity shrinks (see Figure \ref{f2}).  
  Interestingly, the weakly twisted lines enter the large flux tube,  while the 
  short twisted lines marked by the arrow in Figure \ref{f3} are left 
  where the strongly twisted lines existed at $t$=0.0. The contour of the  
  strong twists ($T_n$=1.0) is plotted on the solar surface at $t$=0.0 and 
  $t$=12.5. At $t$=0.0 the footpoint of the red field lines is rooted in the 
  region surrounded by the contour of $T_n$=1.0, while at $t$=12.5 this contour 
  surrounds the footpoint of the blue field lines.

  This result indicates that the twisted lines that were initially weak convert 
  into the large flux tube having strongly twisted lines with more than a 
  one-turn twist. This process might be explained by the tether-cutting 
  reconnection between the eruptive strongly twisted lines in red and ambient 
  weakly twisted lines in blue, because the weakly twisted lines can receive 
  magnetic helicity from the strongly twisted lines via reconnection processes. 
  Therefore, the strongly twisted region on the central positive polarity moves 
  the southern region occupied by the weakly twisted lines as seen in 
  Figure \ref{f2}. 

  Figure \ref{f4} also shows the temporal evolution of the magnetic field lines
  from another angle, plotted in the same format as Figure \ref{f3}, except that
  the evolution of the strongly twisted lines is omitted. The yellow surface 
  corresponds to the critical height of the torus instability, {\it i.e.}, a 
  decay index of $n$=1.5 where $n$ is defined by 
  \begin{equation}
  n(z) = - \frac{z}{|\vec{B}|}\frac{\partial |\vec{B}|}{\partial z},
  \label{EQ_DI}
  \end{equation}
  where, the decay index is calculated based on the potential field. 
  Because the decay index is derived from the external poloidal field sustaining 
  the hoop force by the flux tube(\citealt{2007AN....328..743T}), it is 
  difficult to separate the external field from the NLFFF, following 
  \cite{2007ApJ...668.1232F} and  \cite{2010ApJ...708..314A}. This result 
  clearly shows that the weakly twisted lines reform into large and strongly 
  twisted lines through reconnection before reaching the critical height of 
  the torus instability. After this, the top of the field lines exceed the 
  critical height and would further grow into the CME.

  We carry out a more detailed quantitative analysis in terms of the 
  conservation of magnetic helicity. Figure \ref{f5}(a) displays the 
  distribution of the magnetic twist at $t$=0.0 with the contours of $T_n$=1.0 
  at $t$=15.0 and $B_z$. Although one footpoint of the large flux tube formed 
  during the flare is observed within a region surrounded by the red line 
  in Figure \ref{f5}(a), the twist value of the field lines at this location at 
  $t$=0.0 is less than 0.375 turns, approximately. The twisted field lines at 
  $t$=0.0 are plotted in Figure \ref{f5}(b), where the red twisted lines have 
  more than a one-turn twist while one of the blue twisted lines, 'WL1', has 
  less than 0.375 turn twist approximately and another blue twisted line, 'WL2' 
  has a twist close to zero. However, very few field lines in 'WL2' have more 
  than a half-turn twist, as shown in Figure \ref{f5}(c). If the tether-cutting 
  reconnection takes place only between the field lines 'WL1' and 'WL2' , then 
  the long twisted lines with more than a one-turn twist are never produced, 
  since by the conservation law of the magnetic helicity. WL1 and WL2 do not 
  supply the twist enough to form such lines. This large flux tube is, 
  therefore, formed through the reconnection between the weakly twisted lines 
  'WL1', and 'WL2', and strongly twisted lines which are a source supplying 
  the magnetic helicity. 

  The field lines at $t$=0.0 and $t$=15.0 traced from the same location on the 
  positive polarity are plotted with $B_z$ distribution and contours of 
  $T_n$=0.5 and 1.0 at $t$=15.0 in Figures \ref{f5}(d) and (e), respectively. 
  As we presented in Figure \ref{f3}, at $t$=15.0 the blue twisted lines become 
  strongly twisted lines, even though they were weakly twisted at $t$=0.0. 
  Eventually, the footpoint of the red field lines dominates at $t$=15.0 at a 
  location marked by the dashed white circle, which is where the blue field 
  lines resided at $t$=0.0. These results also support that magnetic 
  reconnection takes place between the strongly twisted lines and weakly twisted
  lines, WL1 and WL2;consequently, part of the red lines remain at the solar 
  surface at $t$=15.0 as sheared post-flare loops. Thus, some of the magnetic 
  helicity accumulated into the strongly twisted lines at $t$=0.0 is transferred
  into the weakly twisted lines through the reconnection, resulting in the 
  formation of the large strongly twisted lines.

  \subsection{Reconnection Dynamics in the X2.2 Solar Flare}
  We investigate the reconnection process in the early flare phase. Following 
  \cite{2013ApJ...773..128T} and \cite{2014ApJ...788..182I}, we estimate 
  the reconnected field lines by using a spatial variance of the field 
  line connectivity by allowing them only by reconnection. The formulation is 
  defined by the following equation:
   \[
   \delta(\vec{x}_0,t_n)  = |\vec{x}_{1}(\vec{x}_0, t_{n+1})
                                            - \vec{x}_{1}(\vec{x}_0, t_{n})| 
                            \quad (\mbox{$T_n \geqq 0.3$}),
  \]
  where $\vec{x}_{1}(\vec{x}_0, t_{n})$ is the location of one footpoint of
  each field line at time ${\it t_n}$, which is traced from another footpoint
  at $\vec{x}_0$. Eventually, we calculate
  \begin{equation}
    \Delta(\vec{x}_0,t) = \int_{0}^{t} \delta(\vec{x}_0,t_n)dt_n,
    \label{foot}
  \end{equation}
  where the enhanced region in $\Delta(\vec{x}_0, t)$ indicates memories
  in which the reconnection took place dramatically in the twisted lines. 
  Figure \ref{f6}(a) shows the distribution of $\Delta(\vec{x_0},t=2.0)$ 
  over the $B_z$($>$0.01) distribution, which we previously discussed, and 
  this shape is similar to that of an observed two-ribbon flare 
  (\citealt{2014ApJ...788..182I}.) Figure \ref{f6}(b) shows the $\Delta$ 
  map with the twist contours  $T_n$=0.5 and 1.0 over the Figure \ref{f6}(a). 
  From this figure we found that the enhanced region 'T2' is found just outside 
  of the region containing the strongly twisted lines, while the region 'T1' 
  overlaps with their insides. We can find the same profiles on the positive 
  polarities. Figure \ref{f6}(c) shows the 3D magnetic field lines, where the 
  red field lines indicate the strongly twisted lines with more than a one-turn 
  twist and where the orange lines go through an inner part of the region 
  surrounded by the current density $|\vec{J}|$=5.0 contour. In an earlier 
  flare phase the strongly twisted lines with more than a one-turn twist erupt 
  away from the solar surface, beneath which the orange field lines are 
  reconnected, forming the long twisted lines and post-flare loops on the 
  $\Delta$ map. 

  We investigate the relationship between the field lines structure and EUV 
  images taken by {\it SDO} to further understand the reconnection dynamics.
  Figures \ref{f7}(a)-(c) represent the field lines structure at $t$=0, where  
  each colored field line( red, orange, and blue) is traced from the regions, 
  R1, R2, and R3 surrounded by dashed circles in Figure \ref{f6}(a). The 
  intensity contour taken from an AIA 171 {\AA} image is also plotted with 
  these field line structures, where Figures \ref{f7}(a) and (b) show the much 
  earlier flare phase and Figure \ref{f7}(d) exhibits it in middle flare phase. 
  In Figures \ref{f7}(a) and (b), the most enhanced areas are located at E1 and 
  E2 for each time where the small flux tubes are coexisting, {\it e.g.}, we 
  can see the red and orange field lines at E1 and three different field lines 
  at E2. This means that the reconnection might take place among these small 
  flux tubes. Figure \ref{f7}(d) shows the field lines structure $t$=2.0. The 
  field lines traced from the regions R1 and R2 convert into post-flare loops, 
  on the other hand, some of footpoint in blue traced from R3 are rooted into 
  the hook region on another polarity in which the contours of the strongly 
  twisted lines are also observed, as shown in Figure \ref{f6}(b). Another 
  footpoint is rooted into the negative polarity located southeast away from 
  the central area. This location is marked by a solid white circle at each 
  time in Figures \ref{f7}(a)-(c), in which EUV images are enhanced during the 
  flare. These results demonstrate that the location close to the PIL is 
  crowded with some twisted lines, {\it i.e.}, small flux tubes, before the 
  flare, and then the complicated 
  reconnection takes place among them during the flare.

  We also checked the $\Delta$ map and 3D magnetic field at $t$=4.0.
  Figure \ref{f8}(a) shows the $\Delta(\vec{x}_0,t=4.0)$ with the twist contours
  mapped on the bottom surface in the same format as that in Figure \ref{f6}(b),
  and it well resembles the two-ribbon flare structure 
  (see \citealt{2014ApJ...788..182I}). Figure \ref{f8}(b) exhibits the 3D 
  structure of field lines in the same format as Figure \ref{f6}(c), except 
  the distribution of the vertical velocity is plotted on the vertical cross 
  section. Figure \ref{f8}(c) shows the enlarged view of the current layer and 
  the distribution of vertical velocity close to the solar surface. We clearly 
  see the upward and downward flows, which have an origin at the inner side of 
  the current layer, meaning that the reconnection is occurring here. The 
  strongly twisted lines in red are a slightly different from that seen in 
  Figure \ref{f6}(c). We can see that one footpoint is rooted into the negative 
  polarity located southeast away from the central region. This result indicates 
  that the strongly twisted lines also interact with the ambient twisted lines 
  during their ascension, as suggested by Figure \ref{f3}. The orange field lines
  show similar behavior as that seen in Figure \ref{f6}(c). Through reconnection,
  the field lines convert into the long twisted field line, whose one footpoint 
  is rooted into the negative polarity in the southeast, and the post-flare 
  loops. Figure \ref{f8}(b) also clearly shows that the locations of both 
  footpoints of the post-flare loops correspond to those in which the $\Delta$ 
  maps exist. 
  
  \subsection{Comparisons with the Observations }
  \subsubsection{Enhancement of the Tangential Field at Solar Surface}  
  In this section, we compare our simulation results with some observations to 
  evaluate the validity of our reconnection dynamics. 

  \cite{2012ApJ...745L..17W} discovered an enhancement of the tangential magnetic
  fields during the flare, residing in a location close to the PIL on the solar 
  surface. They, as well as \cite{2012ApJ...748...77S}, suggested that this 
  enhancement is strongly related to the reconnection. We check this with the 
  temporal evolution of the tangential fields ($B_t$=$\sqrt{B_x^2 + B_y^2}$) 
  obtained from our simulation. Figure \ref{f9}(a) shows the $B_z$ distribution 
  in the central area in which the tangential fields are measured. 
  Figures \ref{f9}(b) and (c) represent the distribution of the tangential 
  magnetic fields at $t$=0.5, and $t$=15.0, respectively. We clearly see that 
  the tangential fields are strongly enhanced at a location close to the PIL, 
  whose value can reach more than 2000 G. These values and locations are 
  therefore well consistent with those observed by \cite{2012ApJ...745L..17W}. 

  The structure of field lines is depicted in an early flare phase above 
  the enhancement of $B_t$ at the photosphere, at $t$=0.0, 1.5, and 3.0 in 
  Figures \ref{f9}(d), respectively.  At $t$=0.0, just after the MHD relaxation 
  process, the long sheared field lines are formed, and then they convert into 
  the short loops at $t$=1.5. Since these short field lines pass through the 
  strong current region, we can interpret that these correspond to post-flare 
  loops formed through the reconnection. Furthermore, $B_t$ is enhanced under 
  the area in which the post-flare loops reside, even in the early phase at 
  $t$=1.5. At $t$=3.0, we found that $B_t$ is more enhanced and the taller 
  post-flare loops are formed, above which the further new post-flare loops 
  are being produced and piled up through sequential reconnection. This result 
  implies that the tangential field of the newly formed post-flare loops 
  sufficiently compresses the pre-existing ones below, resulting in the 
  enhancement of the tangential magnetic fields near the solar surface.
 
   In Figure \ref{f9}(e), we show a temporal evolution of an averaged shear 
  angle between the tangential fields in the simulation and that of the potential 
  field, which is measured on $B_t > 0.4$ at the photosphere, surrounded by 
  white dashed square in Figure \ref{f9}(b). The shear angle is defined by
  $\psi = cos^{-1}(\vec{B}_t \cdot \vec{B}_{pt})/|\vec{B}_t||\vec{B}_{pt}|$ 
  where $\vec{B}_t=(B_x,B_y)$, and $\vec{B}_{pt}=(B_{px},B_{py})$ (tangential 
  components of the potential field). Following this result, the averaged shear 
  angle $<\psi>$ is decreasing, {\it i.e.}, relaxing toward the potential field. 
  These results indicate that the magnetic fields become overall less energetic 
  as shown in \cite{2007ApJ...655..606S} even though $B_t$ is enhanced with time.
  
  \subsubsection{Enhancement of the normal current density at the Solar Surface}
  \cite{2014ApJ...788...60J} reported an enhancement of the normal current 
  density ($J_z$) on the solar surface during the flare, particularly at the 
  negative polarity in the central area (Region S- shown in Figure 3 in their
  paper). We also check the temporal evolution of the negative current density 
  flux to compare with their observational result measured in the region S-. 
  Figure \ref{f10}(a) shows a distribution of $J_z$ at $t$=0.5 and orange line 
  corresponds to a contour of $J_z$=-35 at which the $J_z$ is strongly enhanced. 
  The normal current flux is measured in an area surrounded by red dashed square 
  in Figure \ref{f10}(a) which almost touches to the enhanced region of $J_z$ 
  and close to the region S- set in \cite{2014ApJ...788...60J}. A temporal 
  evolution of the negative normal current flux $\int |-J_z|dS$ is plotted in 
  Figure \ref{f10}(b). Our simulation results also shows its sudden enhancement, 
  but its value decreases 
  gradually after. This is different from the results found by 
  \cite{2014ApJ...788...60J} who observe a saturation of the normal current 
  density. Figure \ref{f10}(c) shows the temporal evolution of the contour for 
  $J_z=-35$. This contour is moving away from the PIL as time progresses as shown
  in \cite{2012A&A...543A.110A}, and the size becomes smaller as speculated in 
  Figure \ref{f10}(a).  
  
   We also checked the relationship between the enhanced current density and 
  field line structure. The structure of field lines at $t$=3.5 passing through 
  the inside of the current layer $|\vec{J}|=10$ is plotted in 
  Figure \ref{f10}(d). The one footpoint of these lines is rooted in the 
  negative current region, where the current density is strongly enhanced.  
  An insert exhibits the top view of the $J_z$ map at $t$=3.5 where the arrow 
  points out the most enhanced region of the negative current density. 
  We further calculate the norm to elucidate the property of these field lines. 
  The norm is defined by following equation equation 
  (\citealt{1996A&A...308..643D}.)  
  \begin{equation}
  \displaystyle N(x,y) = \sqrt{ 
                         \sum_{i=1,2}
                         \left[ 
                           \left( \frac{\partial X_i}{\partial x} \right)^2 
                       +   \left( \frac{\partial X_i}{\partial y} \right)^2 
                         \right]
                              }, 
 \label{qsl}
 \end{equation} 
   where $(X_1, X_2)$ is the relative distance corresponding to 
  $(x'-x'', y'-y'')$. $(x',y')$ and $(x'',y'')$ are the positions of the end 
  points of the field lines whose starting points are two adjacent grid points
  located at ($x'_0$, $y'_0$) and ($x''_0$, $y''_0$) on the photospheric 
  surface. This means that the locations of the end points of these field 
  lines, which are traced from these starting points across a large N(x,y) value, 
  may differ greatly. Figure \ref{f10}(e) plots the field lines at $t$=3.5. These
  are the same field lines as in Figure \ref{f10}(d), with a distribution of
  N(x,y) added. We can clearly see that the one footpoint of these field lines 
  is rooting into the enhanced negative current region and is across the N(x,y) 
  marked by Q1. This means that the footpoints of the field lines in a different 
  topology defined by N(x,y) coexist in the enhanced current layer. We further 
  plot the field lines at $t$=5.0 in Figure \ref{f5}(f), when all of field lines 
  traced from the same location in Figure \ref{f10}(e) become post-flare loops. 
  From these results, the field liens plotted in Figure \ref{f10}(d) or 
  Figure \ref{f10}(e) are in a sate just before and after the reconnection, 
  and we found that the normal current density is strongly enhanced at 
  the moment of reconnection, where the footpoint is rooted. These results are 
  consistent with \cite{2013A&A...555A..77J} demonstrating a relationship between
  the quasi separatrix layer (QSL) and current ribbons formed at the photosphere 
  reproduced in their simulation.

  In our simulation, the normal current flux cannot maintain the peaking value 
  in its temporal evolution. This is in contrast to the observation made by 
  \cite{2014ApJ...788...60J}. One possible reason for this difference might 
  be related to our boundary condition arising from the initial NLFFF. The 
  temporal evolution of the tangential fields in our simulation relaxes the shear
  gradually through the photosphere such that it trends toward the potential 
  field or another low energy state gradually so that cannot trace the evolution 
  as seen in the observation. Also the gap between the observation and 
  simulation shown in Figure \ref{f9}(e) might be caused by this problem. On 
  the other hand, following \cite{2013A&A...555A..77J},  the current ribbons are 
  linked with the current region in the solar corona at which the reconnection 
  taking place, such that numerical diffusion might weaken this link.
  
  \section{Discussion}
  \subsection{Change for the Connectivity in the Strongly Twisted Lines}
  We found that one fooptpoint of the new large flux tube that had the strongly 
  twisted lines and  $T_n>$1.0 and was formed through reconnection during the 
  flare was rooted into the negative polarity located southeast away from the 
  central area as seen in Figures \ref{f7}(d) or \ref{f8}(b). We further check 
  it by comparing it with another AIA images. Figure \ref{f11}(a) exhibits the 
  AIA 171 {\AA} taken by the SDO at 01:48:25 UT on 2011 February 15, 
  corresponding to the onset of the solar flare. The region surrounded by the 
  black square corresponds to the area in which one footpoint of the flux tube 
  was rooted during the flare, while another is rooted into the central area. 
  The temporal evolutions of this area are shown in Figures \ref{f11}(b)-(e). 
  As time progresses during the flare, we can see that the dimming region is 
  growing. Figure \ref{f12}(a) shows another AIA image, taken during the middle 
  flare phase and superimposed with $B_z$ contours. The black dashed circle 
  corresponds to the region surrounded by the square marked in 
  Figure \ref{f11}(a). The strongly twisted field lines at $t$=15.0 are plotted 
  in Figure \ref{f12}(b), where these field lines are traced from the inside of 
  the contour for the twist $T_n$=1.0 at $t$=15.0 in white on the positive 
  polarity. We can see that another footpoint is rooted into the dimming region 
  at the negative polarity, which is marked by the dashed circle in 
  Figure \ref{f12}(a).
  
  Figures \ref{f12}(c) and (d) map the distribution of the open-closed field 
  lines obtained from the NLFFF before and after the flare, respectively, with 
  the field lines traced by the blue solid circle. At the region marked by the 
  blue circle in both panels, the open magnetic field lines going through the 
  side or top boundaries of this field of view (FOV) dominate before the flare, 
  while the closed field lines with both footpoints are anchored within this FOV  
  dominate after the flare. Since this process implies the reconnection between 
  the open and closed field lines and produces the large flux tube as shown in 
  Figure \ref{f3}, this might support the dynamics obtained from our MHD 
  simulation.   

  \subsection{A Relationship between 3D Dynamics and Enhancement of the 
             Tangential Fields at the Solar Surface}
  In section 3.4.1, we saw an enhancement of the tangential magnetic fields at 
  the solar surface that reside close to the PIL. In this section, we explore a 
  relationship between this enhancement and the structure seen in the 3D volume. 
  We first plot the temporal evolution of the maximum value $B_{t_{max}}$ and 
  summation ($B_{t_{sum}}$ (=$\sum B_t$)), for which the $B_t>$0.8 is counted, 
  of the tangential magnetic fields in Figure \ref{f13}(a). These values first 
  increase during the initial flare phase (from $t$=0 to $t$=5.5), where 
  the increasing ratio is 38$\%$ measured in $B_{t_{max}}$, but these 
  values saturate approximately in the late phase, as seen by 
  \cite{2012ApJ...745L..17W}. This transition might be explained by different 3D 
  dynamics in the early and late flare phases. We therefore check the 3D magnetic
   structure to explain this transition.

  Figure \ref{f13}(b) shows the 3D view of the magnetic field at $t$=3.5, 5.0, 
  and 6.5, respectively, with current density contours $|\vec{J}|=5.0$ and 
  isosurfaces corresponding to the critical height of the torus instability. At 
  $t$=3.5, when $B_{t_{max}}$ and $B_{t_{sum}}$ are increasing, the weakly 
  twisted lines in blue which become the strongly twisted lines at $t$=6.5 are 
  passing through the inside of the strong current region. This means that the  
  large flux tube is forming in this phase through reconnection. At $t$=5.0 the 
  large flux tube has been formed and its top is close to the location of the
  threshold of the torus instability. The profile of $B_{t_{max}}$ or 
  $B_{t_{sum}}$ are just about to begin their saturation phase. Eventually, 
  the large flux tube exceeds the threshold of the torus instability;then, 
  $B_{t_{max}}$ or $B_{t_{sum}}$ remain in the saturation phase. These results 
  imply the profile of $B_t$ shows two different picture in the formation 
  of the large flux tube and its ascending phase over the threshold of the 
  torus instability. The different 3D dynamics in the solar corona might be 
  shown at the solar surface in the form of the enhancement of $B_t$.   

  Essentially, this would be controlled by the reconnection in the solar 
  corona. In the earlier flare phase, the reconnection takes place in the lower 
  corona, meaning that the magnetic field is reconnected effectively and can 
  enhance the tangential magnetic field at the photosphere because the strong 
  magnetic flux density exists there. On the other hand, at a later phase, the 
  magnetic reconnection point shifts toward the higher coronal region associated 
  with the rise of the flux tube, so the reconnection is not working effectively 
  relative to the one in the lower corona, which might put the brake on the 
  enhancement of the tangential fields at the photosphere. These more detailed
  and quantitative analyses not presented here are left as future work.

  \subsection{A Transition from Flare to CME}
  Figure \ref{f14} shows a summary of the dynamics for the X2.2-class solar flare
  in AR11158 based on our simulation results. The onset of the solar flare 
  starts in phase (i) where the strongly twisted lines are produced via 
  tether-cutting reconnection of the twisted lines formed in the NLFFF. 
  Consequently, the strongly twisted lines erupt away from the solar surface, and 
  the sheared two-ribbon flares are appearing and new large flux tube is being 
  formed in phase (ii). The CME onset starts in phase (iii), where the newly 
  formed large flux tube is ascending upward over the threshold of the torus 
  instability.

  \cite{2006ApJ...645..742I} suggested an existence of a critical height above
  which a flux tube cannot recover to the equilibrium state. They concluded
  that the CME formation due to the eruption in the flux tube is not enough to 
  surpass the critical twist due to the kink instability;however, it is important
  to couple this  with the loss of equilibrium such that the flux tube can grow 
  over a certain height. Reaching the critical height of the torus instability 
  would equate to the point where equilibrium is lost. In this study, the 
  tether-cutting reconnection plays an important role as driver in the lower 
  corona at the initial phase of flare-CME dynamics, rather than the kink 
  instability. However, a scenario of the flare-CME relationship is basically 
  the same as in our previous study \cite{2006ApJ...645..742I}.

  \section{Summary}
   We investigated the dynamics of the magnetic field during the X2.2-class 
  solar flare taking place in AR11158, in particular focusing on the large 
  flux tube formation and the related phenomena associated with the flares, 
  eventually suggesting a relationship between flares and CMEs.
  \begin{enumerate}
  \item
  We first reconstruct the NLFFF based on the photospheric magnetic field prior 
  to the X2.2-class solar flare in AR11158 on 2011 February 15. This NLFFF never 
  showed the dramatic dynamics as seen in observations because it is stable even 
  against small perturbations (see \citealt{2014ApJ...788..182I}), {\it i.e.}, 
  it is stable against the ideal MHD instability such as like kink and torus 
  instabilities. The NLFFF, therefore, needs to be excited into higher energy 
  levels to cause the huge flares. 

  \item
  We further performed the MHD relaxation, {\it i.e.}, the MHD simulation with 
  the fixed boundary condition where the three components of the  magnetic field
  are fixed at the bottom surface, using the NLFFF as an initial condition. 
  An anomalous resistivity is applied in this calculation and as a result, the 
  strongly twisted lines are produced through the reconnection between twisted 
  lines taking place in the strong current region formed in the NLFFF. Because 
  this magnetic configuration cannot keep the equilibrium state, the strongly
  twisted lines successfully showed the eruption due to a loss of equilibrium. 
  In fact, the same kinds of trigger processes are needed, rather than the  
  anomalous resistivity, to produce the strongly twisted lines through the 
  tether-cutting reconnection and bring them up to a critical height of the 
  torus instability. We here point out some pioneering works about the triggering
  process of the solar flare: Flux cancellation on the photosphere
  (\citealt{1989ApJ...343..971V}; 
   \citealt{2011A&A...526A...2G};
   \citealt{2010ApJ...717L..26A}; 
   \citealt{2011ApJ...742L..27A}),  
  emerging magnetic flux through the photosphere 
  (\citealt{1977ApJ...216..123H};
   \citealt{1995JGR...100.3355F};
   \citealt{2000ApJ...545..524C};
   \citealt{2012NatPh...8..845R};
   \citealt{2012ApJ...760...31K};
   \citealt{2013ApJ...773..128T};
   \citealt{2013ApJ...778...48B}).
  These processes should be taken in account in our simulation as important 
  future work so that our model does not depend on the anomalous resistivity 
  model.

  \item
  Although the strongly twisted lines with more than a one-turn twist launch 
  away from the solar surface, they further reconnected with the ambient weakly 
  twisted lines. Consequently, the new large flux tube is formed. This result 
  shows that the large flux tube growing into the CME is formed through the 
  tether-cutting reconnection between the strongly and weakly twisted lines.
  The tether-cutting reconnection, therefore, plays an important role in the 
  triggering process of the CME, {\it i.e.}, the formation process of the large 
  flux tube and overcoming the threshold of the torus instability, as well as 
  the triggering process of the flare, {\it i.e.}, the formation of the strongly 
  twisted lines in the NLFFF driving the initial eruption. Our results suggested 
  that the flare and CME are connected via these two scenarios.

  \item
  We investigated a relationship between the distribution of $\Delta$ defined in 
  equation (\ref{foot}), corresponding to the connectivity change of a twisted 
  field line, and the distribution of the twisted lines. This $\Delta$ 
  distribution well resembles the distribution of the observed two-ribbon flare 
  in the earlier flare phase. Most of the strong $\Delta$ region is found just 
  outside of the region of the strongly twisted magnetic field. On the other 
  hand, we found that the footpoint of the long twisted lines is anchored into 
  the hook area formed in the $\Delta$ distribution surrounding the footpoint of 
  the eruptive strongly twisted lines. This scenario would be similar to the 3D 
  flare model in \cite{2012A&A...543A.110A} and \cite{2013A&A...555A..77J}. 
  However, following this point, the eruptive strongly twisted lines showed 
  complicated behaviors, in particular, the reconnection between the strongly 
  and weakly twisted lines resulting from the complex magnetic topology of AR 
  11158, through which the large flux tube is formed as mentioned above.

  \item
  In our simulation, we observed a portion of the tangential magnetic field, 
  $B_t$ at the solar surface residing close to the PIL in the earlier flare 
  phase, whose maximum value was recorded to be more than 2000 G. During this 
  phase, the dynamics of the twisted lines are observed within a region less 
  than the critical height of the torus instability in which the large flux 
  tube formation is beginning. We further found that the post-flare loops 
  already exist and cover the region on the photosphere where $B_t$ is enhanced,  
  even though the shear angle between the tangential components in the 
  simulation and that of the potential fields decreases there gradually with 
  time. Therefore, this enhancement is inferred due to be piled up 
  by the new post-flare loops produced by the sequential reconnection, {\it i.e.},
  $B_t$ on the photosphere is compressed by them and then enhanced. After the 
  large flux tube forms and exceeds a critical height of the torus instability, 
  the value of $B_t$ approximately reaches saturation. The maximum value 
  increases 38$\%$ during the initial flare phase while 
  \cite{2012ApJ...745L..17W} reports the average value at the selected region 
  increases by $\sim$ 30$\%$. These results indicate the different profile of 
  $B_t$ on the photosphere, such that increasing or saturating behaviors might 
  be projected by the different 3D dynamics observed in the solar corona.
  On the other hand, although \cite{2012ApJ...745L..17W} and
  \cite{2012ApJ...748...77S} pointed out the tangential fields become better
  aligned with the PIL during the flare, these 'more parallel field lines' 
  appearing in the lowest area were not reproduced in our simulation.

  \item  
  We also found the enhancement of $J_z$ on the solar surface in our MHD 
  simulation, which is consistent with the analysis of the observed result by 
  \cite{2014ApJ...788...60J}. This enhancement is observed in a region where a 
  footpoint of the field lines in which reconnection is taking place in the 
  strong current region is rooted, as pointed out by \cite{2013A&A...555A..77J}.
  On the other hand, our MHD simulation showed that the value of $J_z$ is 
  decreasing gradually as time passes after its enhancement, which is 
  different from the observed picture. There are still some gaps which remain 
  between the observations and simulation.
  \end{enumerate}

  In this paper we investigate the dynamics of magnetic fields during the 
  X2.2-class flare taking place in AR11158, focusing in particular on the 
  transition dynamics from the solar flare to the CME. Unfortunately, we cannot 
  trace the evolution of the flux tube for a long enough time within the limited 
  volume of the simulation, {\it i.e.}, we cannot conclude whether or not the 
  large flux tube shown in our MHD simulation is in its final form. 
  \cite{2014ApJ...793...53H} reported a very highly twist flux tube 
  traveling an interplanetary, up to 5 turns per AU, which decreases toward the 
  edge. At least, our NLFFF or flux tube just launching from the solar surface 
  whose twist values are concentrated mostly in a range from one turn to 1.5 
  turns cannot explain this highly twist. \cite{2010ApJ...710..456Y} also made 
  an effort to clear the gap of twist between measured at the solar surface and 
  interplanetary. But the problem has been left yet. A large-scale simulation 
  covering the interplanetary space would be important to fill this gap.
 
\begin{figure}
  \epsscale{.9}
  \plotone{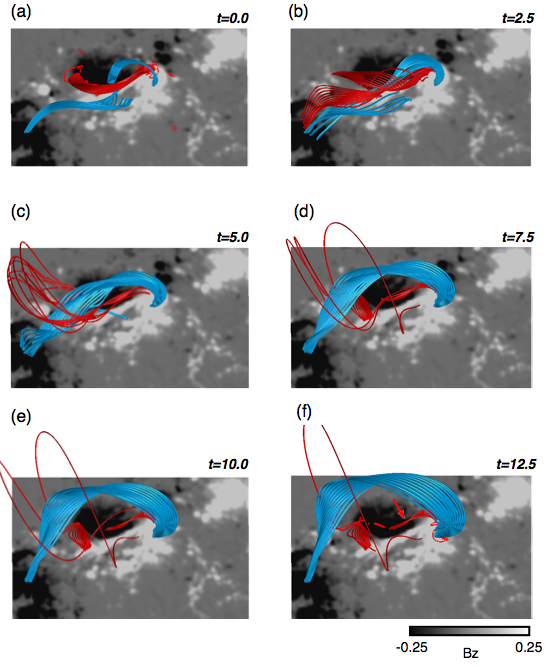}
  \caption{
           Temporal evolution of the dynamics of the selected magnetic field 
           lines with $B_z$ distribution. The red and blue lines represent the 
           field lines more and less than a one-turn twist, respectively, at 
           an initial time $t$=0.0. Note that most of the blue lines have weak 
           twist less than 0.35 turns. The red arrow indicates sheared-post 
           flare loops. The contour of $T_n$=1.0 is plotted at surface in red 
           at $t$=0.0 and 12.5.
          }
  \label{f3}
  \end{figure}

\begin{figure}
  \epsscale{.9}
  \plotone{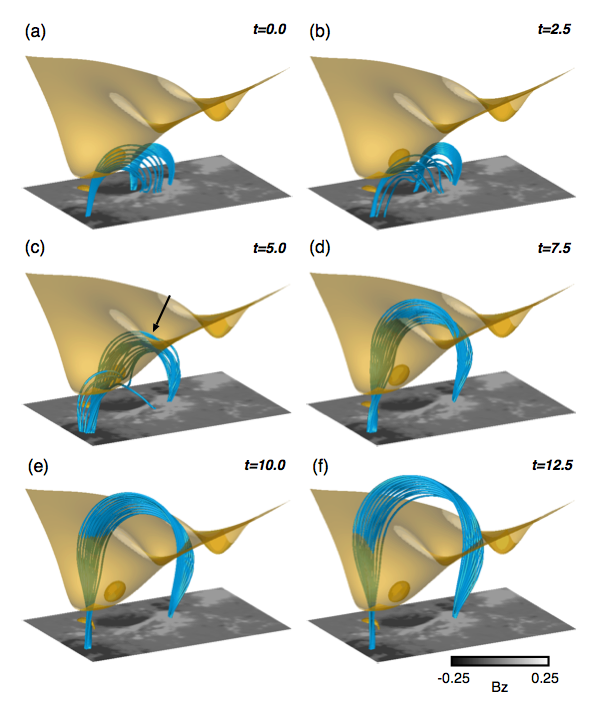}
  \caption{
           Temporal evolution of the dynamics of the magnetic field lines
           from side view of Figure \ref{f3}. The field lines format is same in 
           Figure \ref{f3}, except the temporal evolution of the initial strongly 
           twisted lines with more than a one-turn twist is not plotted. The 
           yellow surface corresponds to a critical height of the torus 
           instability, {\it i.e}, the isosurface with decay index $n$=1.5
           where $n$ is defined by equation (\ref{EQ_DI}). Arrow points towards 
           the intersection of the magnetic field lines and the threshold of the 
           instability.
           }
  \label{f4}
  \end{figure}

\begin{figure}
  \epsscale{1.}
  \plotone{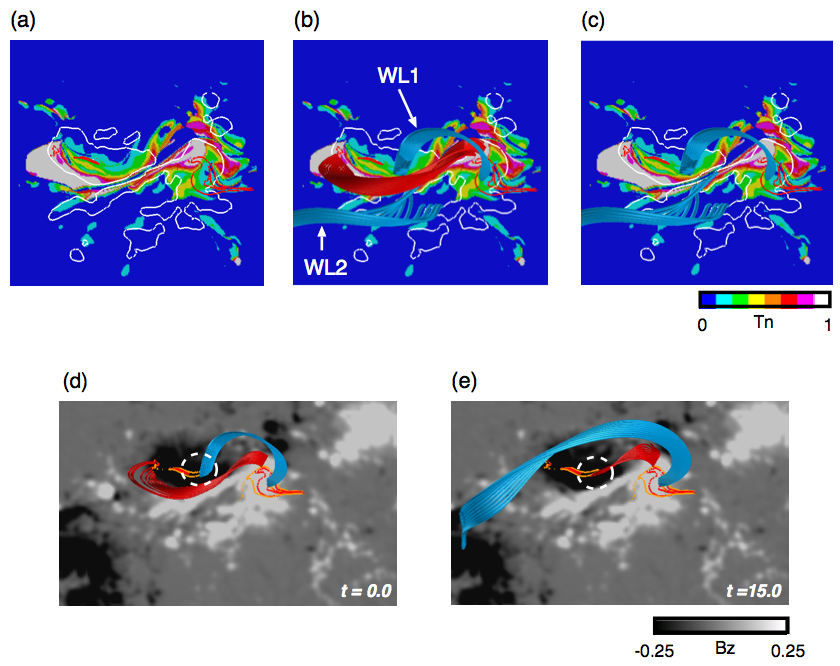}
  \caption{
           (a) The distribution of the twist at $t$=0.0 with contours of
               $T_n$=1.0 at $t$=15.0 in red  and $|B_z|$=0.25 in white.
           (b) The field lines at $t$=0 are plotted over (a). The format
               of these field lines are same as Figure $\ref{f3}$.
           (c) The same field lines at $t$=0.0 are plotted over (a) except the
               strongly twisted lines in red are not plotted.
           (d) The twisted lines at $t$=0.0 are plotted in red and blue, traced 
               from the positive polarity. The contours of twist $T_n$=0.5, and 
               1.0 at $t$=15.0 are plotted in red and orange, respectively, over 
               the $B_z$ distribution where the twisted lines in blue are started
               from the inside region surrounded by the contour $T_n$=1.0 at 
               $t$=15.0.
           (e) The same format with (d) except the field lines are plotted
               at $t$=15.0.
          }
  \label{f5}
  \end{figure}

\begin{figure}
  \epsscale{.9}
  \plotone{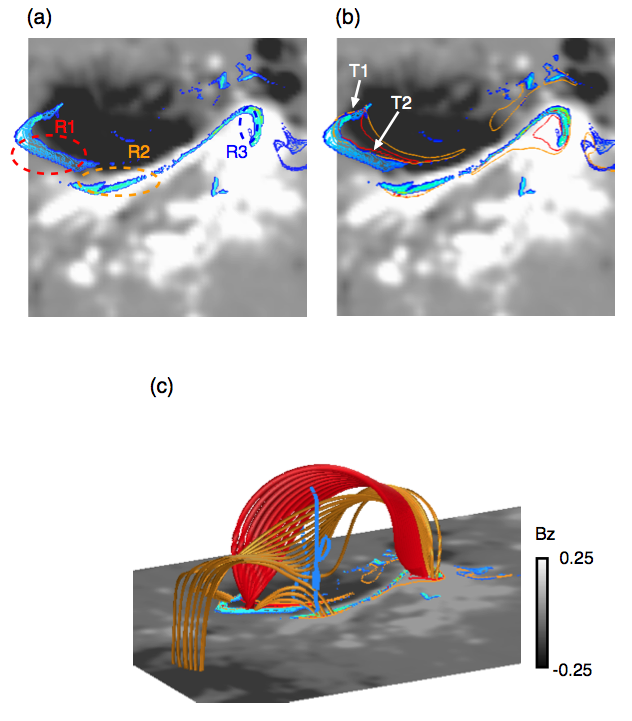}
  \caption{
           (a) $\Delta(\vec{x}_0,2.0)$ is mapped on the $B_z$ distribution 
               where $\Delta(\vec{x}_0,2.0)$ is measured at $|B_z|>0.01$ and 
               $T_n \geqq$0.3.
           (b) The contours of the twist $T_n$ = 0.5 and 1.0 are plotted in
               the orange and red lines with $\Delta$ map over the $B_z$ 
               distribution.
           (c) The 3D magnetic field line structure is plotted over (b). The 
               red lines represent the strongly twisted lines with more than 
               a one-turn twist, while the orange field lines are passing 
               through the region surrounded by the blue line, which is a 
               contour of the current density ($\vec{|J|}$=5.0).
          }
  \label{f6}
  \end{figure}

\begin{figure}
  \epsscale{1.}
  \plotone{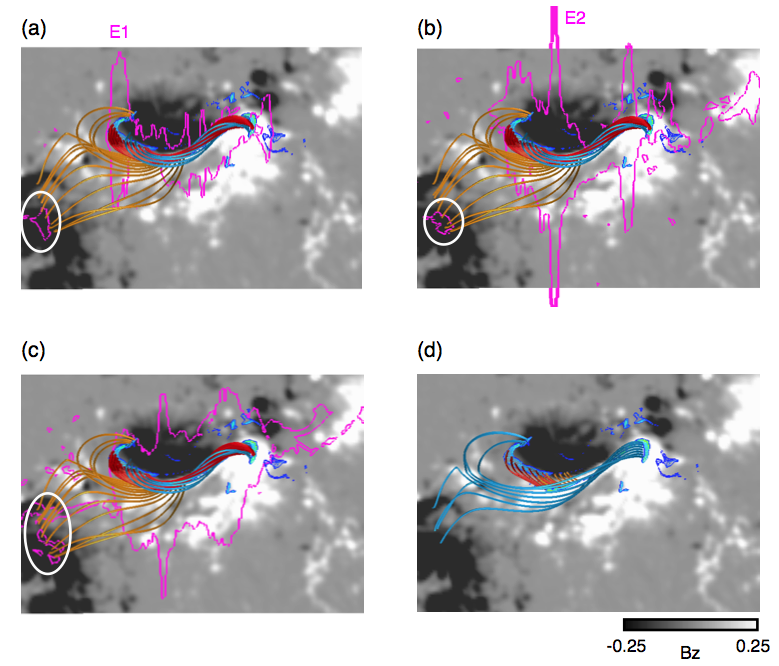}
  \caption{ 
           (a)-(c) The field lines at $t$=0 are plotted with the contour of AIA 
            intensity and $\Delta(\vec{x}_0,2.0)$ over the $B_z$ distribution.
            Each colored field line is traced from each colored region, 
            R1, R2, and R3 marked in Figure \ref{f6}(a). The purple lines 
            correspond to the AIA intensity $log(I)$=9.0 at 01:50:25 UT, 
            01:52:49 UT and 02:04:49 UT from (a) to (c) where the first two
            correspond to the much earlier flare phase and the latter is 
            in the middle flare phase.   
           (d) Each colored field line at $t$=2.0 is plotted over (a), but AIA 
            intensity is not plotted.
          }
  \label{f7}
  \end{figure}

\begin{figure}
  \epsscale{.9}
  \plotone{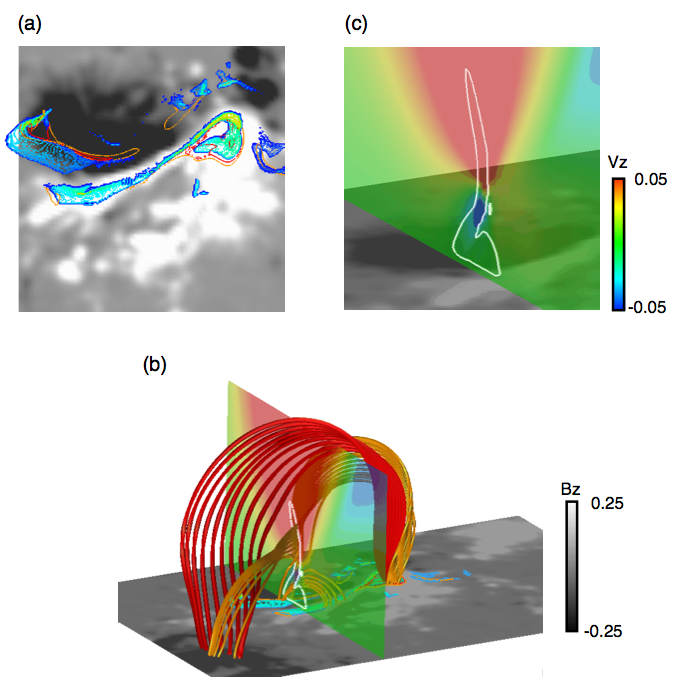}
  \caption{
          (a) $\Delta(\vec{x}_0,4.0)$ is mapped on the $B_z$ distribution at
              the solar surface with the twist contours, $T_n=$0.5 and 
              $T_n$=1.0 in orange and red, respectively.
          (b) The field lines structure at $t=$4.0. The figure format is the 
              same as that in Figure \ref{f6}(c), except the contour of the 
              current density $|\vec{J}|$=5.0 is plotted in white, and the 
              vertical velocity distribution is plotted on the vertical cross 
              section.
          (c) The vertical velocity close to the strong current region is 
              plotted with the current density contour $\vec{|J|}$=5.0.
          }
  \label{f8}
  \end{figure}

\begin{figure}
  \epsscale{.9}
  \plotone{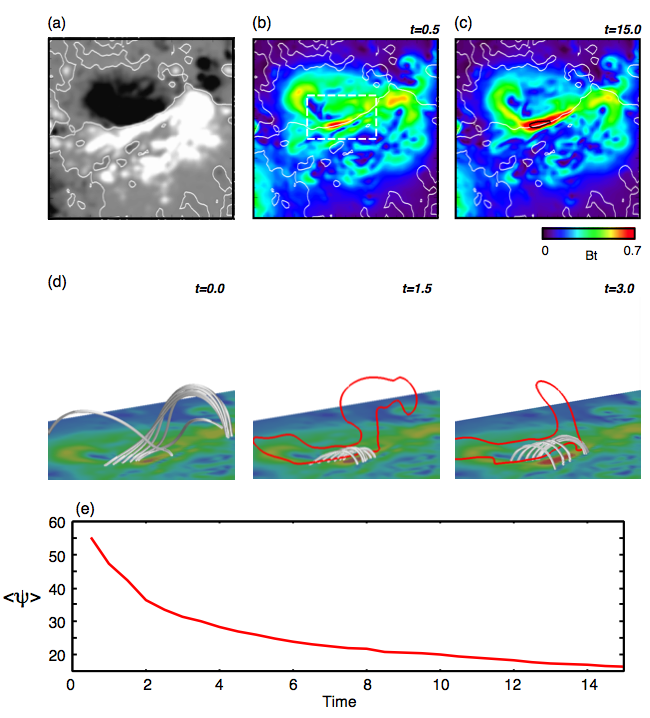}
  \caption{
          (a) $B_z$ distribution with a polarity inversion line in white is 
              plotted in the range of 0.33$<$x$<$0.63 and 0.37$<$y$<$0.67.
          (b)-(c) The distribution of the tangential components of the 
              magnetic field $B_t$=$\sqrt{B_x^2 + B_y^2}$ at $t$=0.5 and 
              $t$=15.0.
          (d) The structure of field lines in the early flare phase is 
              plotted in gray. The field lines shown in (e) and (f) are plotted 
              passing through the inside of the current contour $|\vec{J}|$=10. 
              The field lines in (d) are traced from same points in (e). The 
              bottom surface represents the distribution of $B_t$, which is in 
              the same format as (b) and (c). 
          (e) A temporal evolution of an averaged shear angle between the 
              tangential fields in the simulation and that of the potential 
              fields, measured on $B_t >$ 0.4 at the solar surface, surrounded 
              by white dashed square(0.433$<$x$<$0.533 and 0.50$<$y$<$0.567). 
              depicted in (a). The unit is degree.
          }
  \label{f9}
  \end{figure}

\begin{figure}
  \epsscale{1.}
  \plotone{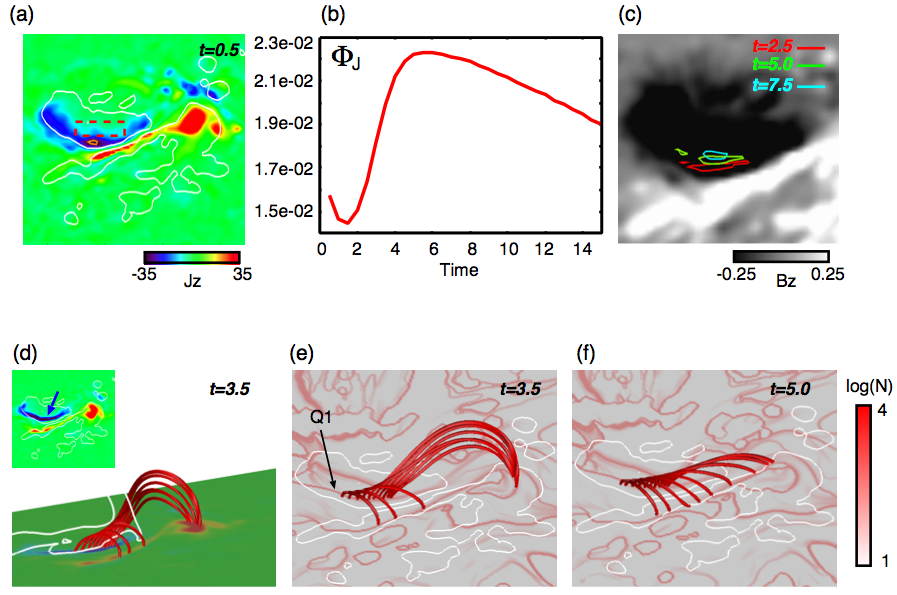}
  \caption{
           (a) A distribution of $J_z$ with contour of $|B_z|$=0.25 in white and 
               $J_z$=$-$35 in orange at $t$=0.5. 
           (b) Temporal evolution of the negative normal current flux, defined 
               by $\int |-J_z|dS$, calculated in a region surrounded by red  
               dashed square (0.433$<$x$<$0.49 and 0.533$<$y$<$0.567) marked in 
               (a).
           (c) Temporal evolution of the contour for the $J_z$=$-$35, at 
               $t$=2.5(in red), 5.0(in green), and 7.5(in aqua), respectively, 
               on $B_z$ distribution. 
           (d) The field lines are plotted in red, passing through the inside 
               region surrounded by the contour of $|\vec{J}|$ = 10 over $J_z$ 
               distributions. Insert is the top view of the $J_z$ map where 
               the arrow points out the most enhanced region.
           (e)-(f) The field lines are plotted in the same format as (c) and are 
                   plotted with a distribution $N(x,y)$ defined in equation 
                   (\ref{qsl}) at $t$=3.5 and $t$=5.0, respectively. The white 
                   contour represents the $|B_z|$=0.25.   
          }
  \label{f10}
  \end{figure}

\begin{figure}
  \epsscale{1.}
  \plotone{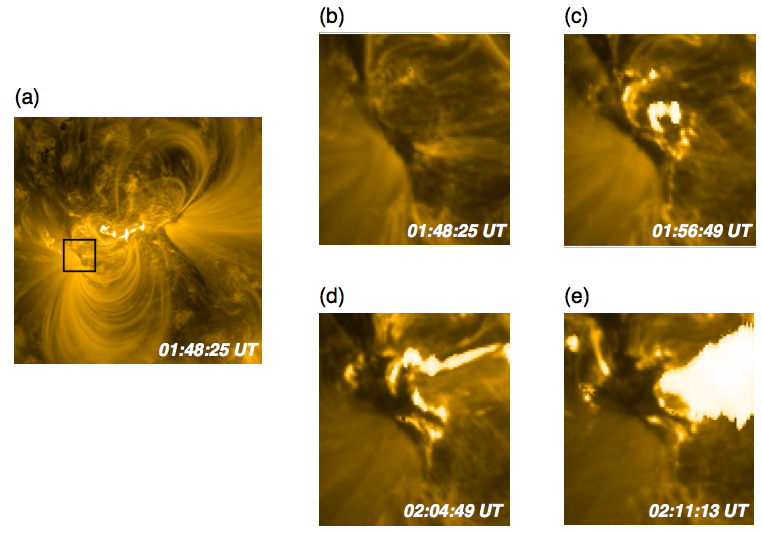}
  \caption{
           (a) EUV image taken from AIA 171 {\AA} at 01:48:25 UT in the range 
               of 216 $\times$ 216 $Mm^2$, corresponding to an area in which the 
               NLFFF or MHD simulation is performed.
           (b)-(e) EUV images taken from AIA 171 {\AA} from $t$= 01:48:25 to 
               $t$=02:11:13 during the flare, in the range of a region surrounded 
               by the black square in (a).                  
          }
  \label{f11}
  \end{figure}

\begin{figure}
  \epsscale{1.}
  \plotone{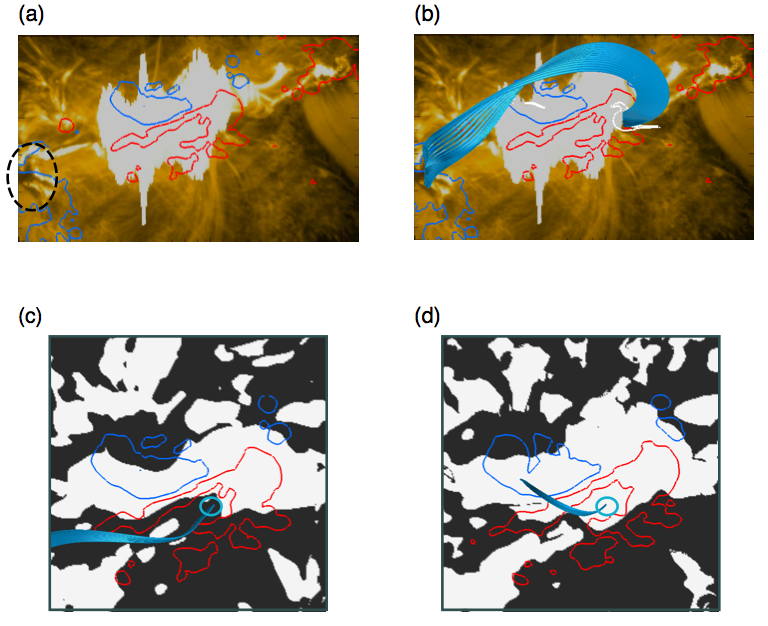}
  \caption{
           (a) EUV image taken by AIA 171 {\AA} in the range of 0.25$<$x$<$0.78, 
               0.33$<$y$<$0.67 at 02:04:49UT. Red and blue lines correspond to 
               the contours of $B_z$=0.25, and -0.25, respectively.
           (b) Field lines with more than a one-turn twist at $t$=15.0, 
               {\it i.e.}, flux tube, are plotted over (a). White line 
               corresponds to the contour of $T_n$=1.0 at $t$=15.0 where the 
               field lines are started from the inside of this region.
           (c) Map of open-closed field lines for the NLFFF on February 15 
               00:00UT corresponding to the 2 hours before the flare. White 
               regions are dominated by the closed field lines where both 
               footpoints are anchored in this field of view (FOV)
               (0.38$<$x$<$0.62, 0.4$<$y$<$0.63), while the black regions are 
               dominated by the open magnetic field whose one footpoint goes 
               through the boundary of this FOV.
               The blue lines are field lines traced from the solid circle.
           (d) Map of open-closed field lines for the NLFFF on February 15 
               03:00UT, corresponding to an hour after the flare. Figure 
               format is same as in (c). 
                }
  \label{f12}
  \end{figure}

\begin{figure}
  \epsscale{1.}
  \plotone{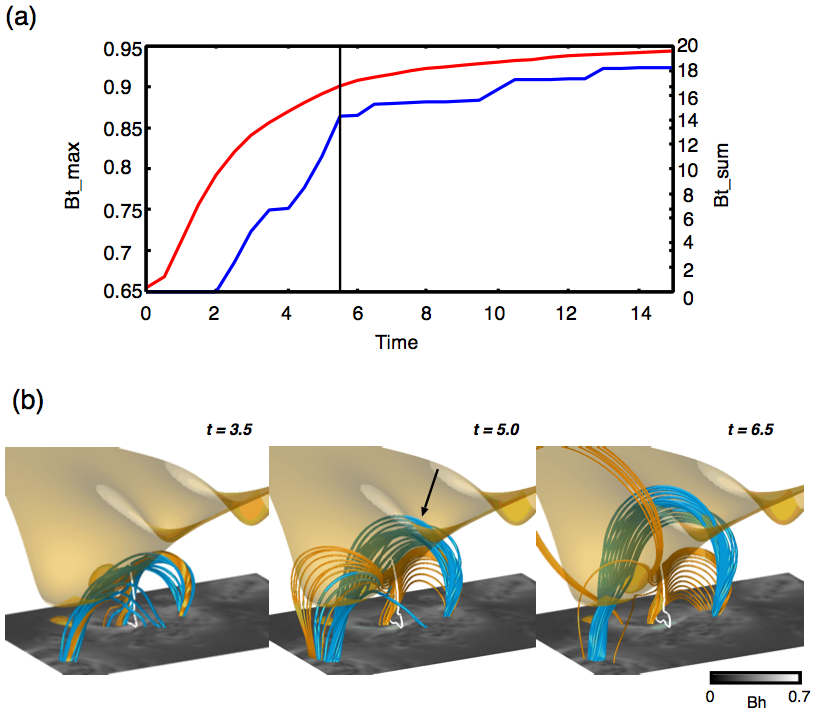}
  \caption{
          (a) A temporal evolution of the maximum value of the tangential 
              components of the magnetic field, $B_{t_{max}}$, and its summation 
              of that, $B_{t_{sum}}$ (=$\sum B_t$) where $B_t$ is counted 
              over 0.8(=2000G), measured in the range of 
              (0.25$L_0$ $<$x$<$ 0.75$L_0$, 0.25$L_0$ $<$y$<$ 0.75$L_0$) are
              plotted in red and blue, respectively. The vertical solid line 
              indicates a location where the profile of the $B_{t_{sum}}$
              suddenly changes from an increasing phase to the saturating phase.
          (b) Temporal evolution of the magnetic field lines with the contour 
              of $|\vec{J}|=5.0$ in white and the isosurface of the threshold 
              of the torus instability. Bottom surface represents the 
              distribution of $B_t$. The blue field lines are drawn in the same 
              format as in Figure \ref{f4}, and orange field lines pass through 
              the region surrounded by the contour of the current density in 
              white. Arrow pointed out the intersection between the magnetic 
              field lines and isosurface.
          }
  \label{f13}
  \end{figure}

\begin{figure}
  \epsscale{1.}
  \plotone{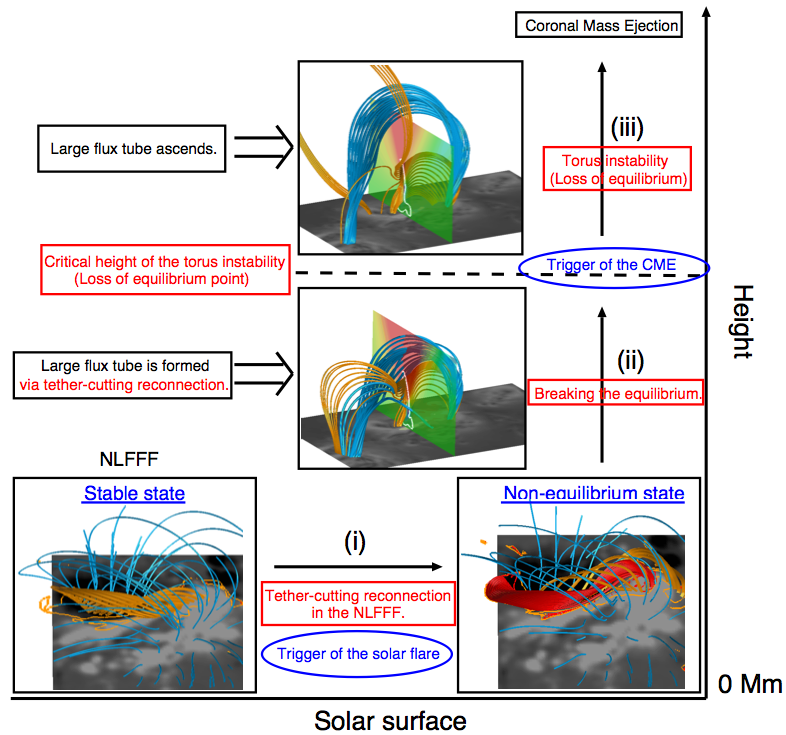}
  \caption{
           Summary of the dynamics in AR11158, based on our NLFFF extrapolations
           and MHD simulations. The upper two insets show the 3D structure of 
           the field lines in the same format as in Figure \ref{f13}(b), except 
           that the vertical velocity is plotted on the vertical cross section, 
           where its top corresponds to the approximate critical height of the 
           torus instability.
          }
  \label{f14}
  \end{figure}

   On the other hand, the twist accumulated in the large flux tube looks weak 
  compared to the initial twisted lines in red in Figure \ref{f3}(a)
  even though the flux tube exceeds a one-turn twist. These would be caused by 
  two problems: one is due to a numerical diffusion 
  (\citealt{2011ApJ...734...53S});
  another is due to the initial NLFFF. In order to settle these problems, a 
  high-resolution simulation is important to suppress the numerical diffusion, 
  {\it i.e.}, suppress the diffusion of the twist accumulated in the flux tube. 
  Furthermore, a higher accurate NLFFF, achieving more close to an equilibrium 
  state, would be also required to avoid release of a shear through the boundary.
  These problems must be settled to complete the data-constrained or -driven 
  simulation for solar flares and CMEs. On the other hand, future observations 
  plan to observe a chromospheric magnetic field ({\it e.g.,} {\it SOLAR-C} 
  mission \citealt{2014SPIE.9143E..1OW}). Because a chromospheric magnetic field 
  is in a lower $\beta$ state than that of the photospheric magnetic field, it 
  would help us definitely construct higher accurate NLFFFs and perform  
  data-constrained or -driven simulations. 
                              
  \acknowledgments
   We thank to anonymous referee for the constructive comments. We are also 
   grateful to Dr.\ Miho Janvier and Prof.\ Kanya Kusano for useful discussions.
   S.\ I.\ was supported by the International Scholarship of Kyung Hee 
   University. This work was supported by the Korea Astronomy and Space 
   Science Institute under the R $\&$ D program (project No.2013-1-600-01) 
   supervised by the Ministry of Science, ICT and Future Planning of the 
   Republic of Korea. G.S.C. was supported by Basic Science Research Program 
   through the National Research Foundation of Korea (NRF) funded by the Ministry 
   of Education (NRF-2013R1A1A2058937). T.\ M.\ was supported by the BK21 
   Plus Program through the National Research Foundation funded by the Ministry 
   of  Education of Korea The computational work was carried out within the 
   computational joint research program at the Solar-Terrestrial Environment 
   Laboratory, Nagoya University. Computer simulation was performed on the 
   Fujitsu PRIMERGY CX250 system of the Information Technology Center, Nagoya  
   University. Data analysis and visualization are performed using resource of 
   the OneSpaceNet in the NICT Science Cloud.

\clearpage

  \clearpage

  \clearpage
  
  \clearpage

  \clearpage

  \clearpage

  \clearpage

  \clearpage

  \clearpage

  \clearpage

  \clearpage

  \clearpage

  \clearpage

  \end{document}